\documentclass[12pt]{article}
\usepackage{amsfonts}
\usepackage{amsmath}
\usepackage{epsfig}

\begin{document}

\begin{flushright}
\end{flushright}
\vspace{20mm}
\begin{center}
\large {\bf CHIRALITY AND SYMMETRY BREAKING IN A DISCRETE INTERNAL SPACE}\\
\mbox{ }\\
\normalsize
\vspace{1.0cm}
{\bf Bodo Lampe} \\              
\vspace{0.3cm}
\vspace{3.0cm}
{\bf Abstract}\\
\end{center} 
In previous papers the permutation group $S_4$ has been suggested as an
ordering scheme for elementary particles, and the appearance of this
finite symmetry group was taken as indication for the existence of a 
discrete inner symmetry space underlying elementary particle interactions. 
Here it is pointed out that a more suitable choice than the tetrahedral group 
$S_4$ is the pyritohedral group $A_4 \times Z_2$ because its vibrational
spectrum exhibits exactly the mass multiplet structure of the 3 fermion
generations. Furthermore it is noted that the same structure can 
also be obtained from a primordial symmetry breaking $S_4 \rightarrow A_4$. 
Since $A_4$ is a chiral group, while $S_4$ is 
achiral, an argument can be given why the chirality of the inner
pyritohedral symmetry leads to parity violation of the weak interactions. 



\newpage

\normalsize



\section{Introduction}

In the left-right symmetric standard model with gauge group 
$U(1)_{B-L}\times SU(3)_c \times SU(2)_L \times SU(2)_R$ \cite{su2su2} 
there are 24 left-handed and 24 right-handed quark and lepton fields 
which including antiparticles amounts to 96 fermionic degrees of freedom, 
i.e. this model has right handed neutrinos as well as 
righthanded weak interactions. 

In recent papers \cite{lampe1,lampe2,lampe3} a new ordering scheme for the observed 
spectrum of quarks and leptons was presented, which relies on the 
structure of the group of permutations $S_4$ of four objects, and a 
mechanism was proposed, how 'germs' of the Standard Model 
interactions emerge from this symmetry. 

In those papers a constituent picture was suggested where quarks and leptons are
assumed to be built from 'tetronic' constituents, whose interchanges generate
the permutation group $S_4$. In the present paper I follow a different
approach which relies on the fact that $S_4$ is not only the symmetry group 
of a regular tetrahedron, but can appear as the point group of several 
cubic lattices\cite{bilbao,johnson}. 
In this approach the inner symmetry space is not continuous
(with a continuous symmetry group) but has instead the discrete 
structure of a 3-dimensional cubic lattice\footnote{It seems 
then natural to assume that not only the internal
symmetry is discrete but that physical space is a lattice, too. 
Although theories with a discrete inner symmetry over a continuous base
manifold have been examined\cite{belavin} they seem to me 
a bit artificial because they 
usually lead to domain walls and other discontinuities. 
Nevertheless, this point may be left open here, because 
for most arguments in this article it is not essential, 
whether physical space is discrete or continous.} 
and the observed quarks and leptons can be interpreted as excitations on this
lattice and characterized by representations of the lattice point group 
$S_4$. 

In the following sections I will discuss in detail the 
nature of these excitations and relate them to the 
Standard Model phenomenology. 
I will further argue that the tetron model is 
is not just a strange observation in the sideways of particle 
physics, but has a more fundamental meaning and may shed light 
on some important issues of high energy physics. Actually, 
in modern particle physics there are several outstanding problems 
which have not been fully understood for many decades: 
they are the parity violation in weak interactions, 
the family structure of quarks and leptons, 
the calculation of their masses and CKM matrix elements 
and the existence of UV-divergences at very small distances. 

In my article I want to analyze these phenomena 
in the light of the tetron model. Tetron interactions 
will be assumed to describe a deeper level of matter 
than the Standard Model, 
somewhere between the Planck scale and the electroweak breaking scale. 

The outline of the paper is as follows: section 2 
describes the role of finite groups in tetron theory 
and shows how the correct multiplet structure of quarks 
and leptons can be obtained. 
In section 3 I elaborate on vibrational modes in 
the inner symmetry crystal (which will be named 'phinons'). 
I will show how the family structure of the Standard 
model fermions naturally arises in this framework. 
In addition I discuss the 
question whether the crystal can be compactified 
to an (inner) 'molecule'. 
Section 4 describes the Standard Model gauge symmetry as a 
phenomenon which arises from phinon-phinon interactions. 
Section 5 deals with a phase transition 
from an achiral to a chiral internal lattice. It will be shown 
that the corresponding symmetry breaking underlies 
the Standard Model Higgs mechanism. 
In section 6 the 7-dimensional cross product is 
introduced as a possible effective interaction 
between phinons, and it is shown that the chirality 
of the inner symmetry lattice induces 
the parity violation as found in the weak interactions. 
Finally, in section 7 an alternative scheme is presented 
which relies on spin waves instead of vibrational 
modes in the inner symmetry crystal. 


\section{The Role of Finite Groups in Ordering 
the Spectrum of Quarks and Leptons} 

While $S_4$ has been discussed extensively in earlier 
papers, in this letter the focus will be on the so-called 
symmetric (or alternating) group $A_4$ which 
is the subgroup of $S_4$ consisting 
of even permutations of 4 objects. It is isomorphic to the group 
of proper rotations of a regular tetrahedron, and is 
therefore a 'chiral' symmetry (in the sense that it does not 
contain any inversions or reflections). 


\begin{table}
\label{taba4}
\begin{center}
\begin{tabular}{|l|c|c|c|c|c|}
\hline
                 & I  & 4$C_3$        & 4$C_3^2$        & 3$C_2$  & transformation behavior \\
                 &    &               &                 &         & of translations, rotations \\
\hline
A                & 1  & 1             & 1               & 1 &              \\
A'               & 1  & $e^{+2i\pi/3}$ & $e^{-2i\pi/3}$ & 1 &              \\
A''              & 1  & $e^{-2i\pi/3}$ & $e^{+2i\pi/3}$ & 1 &              \\
T                & 3  & 0 & 0 & -1                          & $(x,y,z)$, $(R_x,R_y,R_z)$ \\
\hline
no. of atoms unchanged  & 4  & 1 & 1 & 0 & \\
$\times$ T       & 12 & 0 & 0 & 0 & \\    
A+A'+A''+3T      & 12 & 0 & 0 & 0 & \\    
\hline
\end{tabular}
\bigskip
\caption{Character table of the symmetric group $A_4$. 
I, 4$C_3$, 4$C_3^2$ and 3$C_2$ are the classes consisting 
of the identity, 4 rotations by $2\pi/3$ and 
4 rotations by $4\pi/3$ (each seen from a vertex) and 
3 rotations by $\pi$. 
T is the vector representation, which 
acts on a regular tetrahedron in 3 dimensions, and 
A, A', A'' can be traced back to representations 
of the cyclic group $Z_3$ (a subgroup of $A_4$).}
\end{center}
\end{table}

I will start with a simple pedagogical example: consider a tetrahedral 
molecule with 4 identical 'atoms' (one on each corner of the tetrahedron). 
In the first rows of table 1 the characters of $A_4$ are shown 
for the various classes and irreducible representations\cite{griff}. 
In the fifth row for each class the number of atoms is given, 
which are left invariant by the corresponding transformations, 
and these numbers are multiplied in the sixth row 
by the characters of the translational representation T 
(which in the case of $A_4$ agrees with that for rotations, 
c.f. the last column of table 1). 
From the last row of the table one can conclude that 
the representation obtained in this procedure can be 
written as the sum $A+A'+A''+3T$ of irreducible representations.

Why am I doing this? The point is, that it 
is exactly the way phonon modes are classified for a lattice 
with $A_4$ symmetry, in which there are 4 'atoms' in the unit cell. 
The only difference with solid state physics being that 
I will be talking about a lattice and its vibrations which 
exist in a 3-dimensional {\it internal} symmetry space. 
For that reason the corresponding excitations will be called 
{\it phinons}.

If one passes from molecules to crystal structures 
there is a complication about which space group should be 
chosen. For the point group $A_4$, for example, there exist 
6 space groups and their corresponding lattices, 
namely 195-199 corresponding to 
international standard symbols P23, F23, I23, P213, I213\cite{bilbao}. 

As a next step lattices with 8 atoms in the unit cell 
are considered. The procedure is similar as before. 
If, for example, the point group of the lattice 
is the full tetrahedral group $S_4$, 
there are 6 possible space groups (215-220
with international standard symbols P43m, F43m, I43m, P43n, F43c, I43d)
\cite{bilbao}. 
Only for the choices 218, 219 and 220 
one obtains the spectrum of phinon modes of the form
\begin{equation} 
A_1+A_2+2E+3T_1+3T_2
\label{eq7hg}
\end{equation}
which was advocated in earlier papers\cite{lampe1,lampe2,lampe3}. 
In those papers the family structure of 
quarks and leptons was discussed on the basis of $S_4$ and 
expression (\ref{eq7hg}). 

In the case of space group 218(=P43n) with 8 atoms in the unit cell 
one has Wyckoff positions 8e: 
\begin{eqnarray}  
W_1=(x,x,x) \quad W_2=(-x,-x,x)& & W_3=(-x,x,-x) \quad W_4(x,-x,-x) \nonumber \\ 
W_5=(x+y,x+y,x+y) & &  
W_6=(-x+y,-x+y,x+y) \nonumber \\
W_7=(-x+y,x+y,-x+y) & & 
W_8=(x+y,-x+y,-x+y)
\label{eq3uijj}
\end{eqnarray}
where $y=\frac{1}{2}$. These points can be interpreted as 
2 tetrahedrons shifted away from each other by the vector $(y,y,y)$. 

For 219 and 220 one should use the Wyckoff positions 32e and 
16c respectively, and take into account 
that for body centered (BCC=I) and face centered (FCC=F) 
space groups like 220 and 219 one usually does not 
use primitive unit cells but centered cells 
where the number of atoms is multiplied 
by the number of centering vectors (2 for I and 4 for F).
For 219 (F43c) the resulting positions are the same as eq. (\ref{eq3uijj}) 
while for 220 (I43d) they are somewhat more complicated: 
\begin{eqnarray}  
U_1=(x,x,x) & & U_2=(-x+\frac{1}{2},-x,x+\frac{1}{2}) \nonumber \\ 
U_3=(-x,x+\frac{1}{2},-x+\frac{1}{2}) & & U_4(x+\frac{1}{2},-x+\frac{1}{2},-x) \nonumber \\ 
U_5=(x+\frac{1}{4},x+\frac{1}{4},x+\frac{1}{4}) & &  U_6=(-x+\frac{1}{4},-x+\frac{3}{4},x+\frac{3}{4}) \nonumber \\
U_7=(x+\frac{3}{4},-x+\frac{1}{4},-x+\frac{3}{4}) & & U_8=(-x+\frac{3}{4},x+\frac{3}{4},-x+\frac{1}{4}) 
\label{eq3zw44}
\end{eqnarray}

A disadvantage of $S_4$ is that it is an achiral symmetry (in the sense 
that it contains reflections). This will become an issue in 
section 6 where weak parity violation will 
be associated with a chiral inner symmetry. 
Furthermore, the lepton mass spectrum does not exactly fit into the 
representations (\ref{eq7hg}). Muon and 
$\tau$-lepton as well as $\nu_{\mu}$ and $\nu_{\tau}$ would be in the 
same multiplet und retain equal masses. 

Therefore an alternative based on the so-called pyritohedral 
symmetry $A_4\times Z_2$ will now be examined. 
Going from $A_4$ (table 1) to $A_4\times Z_2$ is a simple task
because it amounts to replacing each of 
the representations R of table 1 by a combination of $R_g+R_u$ 
of odd and even vibrations under $Z_2$ and yields a 
multiplet spectrum of 24 vibrational modes: 
\begin{eqnarray} 
A_g(e)+A'_g(\mu) +A''_g(\tau)+
A_u(\nu_{e})+A'_u(\nu_{\mu})+A''_u(\nu_{\tau}) \nonumber \\
+T_g(u)+T'_g(c)+T''_g(t)+ T_u(d)+T'_u(s)+T''_u(b) 
\label{eq8hg}
\end{eqnarray}
I have added in brackets the associated fermion flavor because 
the above combination of phinon modes for symmetry $A_4\times Z_2$ 
and 8 atoms in the unit cell 
will be interpreted as the set of 24 quark and 
lepton states. 
Note that the $Z_2$ factor ($u \leftrightarrow g$ in expression (\ref{eq8hg})) 
corresponds to broken weak isospin and puts the weak isospin partners 
naturally in different multiplets with different masses. 

In general, a crystal with N atoms is composed of N/n primitive 
unit cells, each of which containing n atoms. 
For a d-dimensional crystal the total number of 
vibrational degrees of freedom is $d\times N$ 
which are distributed in $d\times n$ 'modes' or 'branches' 
throughout the first Brillouin zone. 
In the present case 
(d=3, n=8) this amounts to the $3\times 8=24$ modes of 
eq. (\ref{eq8hg}). 
As a consequence there is the following simple memo: 
the number of families corresponds to the 
number of dimensions of the discrete inner symmetry space. 
The number of quarks and leptons within a family 
corresponds to the number of atoms in a unit cell of the 
inner symmetry lattice.

I have gone through the list of all finite point groups and found 
that there is no other group which is able to describe the 
mass multiplets as accurately as $A_4\times Z_2$. 
For example, the existence of a 3 dimensional representation
is an essential requirement, 
if one wants to obtain equal masses for quarks 
with the same color, i.e. $m(q_1)=m(q_2)=m(q_3)$, 
but among the 15 groups of order 24\cite{fini} 
only $A_4\times Z_2$ and $S_4$ have 3 dimensional 
representations.\footnote{The nontrivial groups 
of order 24 are $D_8\times Z_3$ and Dic$_12\times Z_2$, 
$S_3\times K$, Dic$_24$, SL(2,3), $Q_8\times Z_3$, 
$A_4\times Z_2$, $S_4$ and a semidirect product 
of $D_8$ and $Z_3$ (in the notation of ref. \cite{besche}).}

There are in fact 7 crystallographic space groups with 
point group $A_4\times Z_2$: 200-206, 
corresponding to the international standard symbols 
Pm3, Pn3, Fm3, Fd3, Im3, Pa3, Ia3. 
{\it All of them} can have 8 atoms in the 
unit cell and lead to the required spectrum expression (\ref{eq8hg}). 
Furthermore, as shown in section 5, all of them 
can be derived from breaking some larger space 
group with the octahedral group $O_h$ 
as its point group. For convenience at this point I just give the 
Wyckoff positions 8c for space group 205 (Pa3) with 8 atoms 
in the unit cell\cite{bilbao}: 
\begin{eqnarray} 
C_1=(x,x,x) \quad & & C_2=(-x+y,-x,x+y) \nonumber \\ 
C_3=(-x,x+y,-x+y) \quad 
    & & C_4=(x+y,-x+y,-x) \nonumber \\
C_5= (-x,-x,-x) \quad & & C_6=(x+y,x,-x+y) \nonumber \\
C_7=(x,-x+y,x+y) \quad
    & & C_8=(-x+y,x+y,x)  
\label{ew61hg}
\end{eqnarray}
again with $y=\frac{1}{2}$. 

One can get insight in the geometry of this configuration 
(which notably is realized in $\alpha N_2$ nitrogen) 
by considering the limiting cases $|x|\gg |y|$ and $|x|\ll |y|$.
For very large x the system reduces to a cube with full (achiral) 
$O_h$ symmetry whereas for very small x one obtains 
2 tetrahedrons lying above each other with full (achiral) 
$S_4$ symmetry. In section 5 symmetry breaking phase 
transitions will be considered which involve distortions 
of the internal lattice from these two limits to finite values of x, and 
it will be shown that a symmetry breaking of the form 
$S_4 \rightarrow A_4$ or $O_h \rightarrow A_4\times Z_2$ 
may have occured when the universe cooled down after the 
big bang. This symmetry breaking turns out to be the basis 
of the breaking of weak SU(2) in the Standard Model.



\section{Inner Lattice or inner Molecule?}

In the last section we have seen that the procedure 
for ordering the quark lepton spectrum is analogous to that for 
phonons in certain crystals, the only difference being 
that the crystal should lie in a 3-dimensional inner symmetry 
space and the displacements should go into those inner directions. 
For that reason these excitations have been named 'phinons'.

There may be other interpretations of the tetrahedral ordering 
like that the spectrum (\ref{eq8hg}) can be obtained by using e.g. 
rotational instead of translational degrees of freedom, 
c.f the last column of table 1. 
In that case one is led to consider spin models\cite{spinm1,spinm2}, 
and the vector displacements of the atomic vibrations 
are replaced by axial vectors $\vec S=\frac{1}{2}\psi^{\dagger} \vec \tau \psi$ 
which fulfil (internal) angular momentum commutation relations 
and can be considered to be built from an (internal) spinor field $\psi$. 
Phinons are replaced by 'mignons' in that context, 
quasi-particle excitations of the 3-dimensional inner spin vector  
$\vec S$ which obey the dynamics of a Heisenberg Hamiltonian. 
This alternative will be discussed later in sect. 7. 
At the present stage I want to stick to 'phinons' as 
the simplest and most effective solution, which will be seen 
to lead to the correct phenomenology. 

The theory for phinons can be developed 
in analogy with the known results for phonons in solid state 
lattices. If one likes one may assume a situation where 
originally there was a discrete (6+1)-dimensional spacetime 
which by some compactification process splitted into an 
3-dimensional internal space and a remaining (3+1)-dimensional physical 
spacetime so that the notions of time and energy and even 
temperature are defined for the inner 
symmetry lattice as well. 

Consider then an inner crystal for any fixed base point on 
Minkowski space and allow for vibrations given by 
displacements of its atoms from their equilibrium 
position which are assumed to be orthogonal to the base space. 
One can then expand the potential energy function around 
the minimum energy configuration. The leading term is the 
ground state energy $E_0$ and the next-to-leading term is quadratic in the 
displacements $\vec u(l,s)$ of the atoms s in a unit cell l: 
\begin{equation} 
U = E_0 + \frac{1}{2} \sum_{l,s,i,l',s',j} k_{ij}(l,s,l',s') 
u_i(l,s) u_j(l',s')
\label{e667}
\end{equation}
where $i,j=1,2,3$ are vector indices of the internal space 
and $k_{ij} (l,s,l',s')$ is the so called 
force or spring constant matrix\cite{phon}
which by definition relates the forces to the displacements. 
Note that s,s' run from 1 to n and l,l' from 1 to N/n, where 
N is the total number of atoms in the crystal.  

For the calculation of the phonon spectrum there are 
two strategies. One is to use the available computer 
routines\cite{phon,stokes}, and the other to 
transform to normal coordinates and then try approximations. 
For example, 
from the geometrical description given after eq. (\ref{ew61hg}) 
one may deduce an ansatz which simplifies the equations 
very much, namely one is lead to assume that there are 
essentially only two different spring constants: 
one ($k_1$) connecting two atoms within 
a tetrahedron, and the other one ($k_2$) connecting 
the 2 atoms within a $Z_2$ pair. $k_2$ may then also be 
considered the coupling strength between two tetrahedrons. 
In the next step one can try a strong coupling expansion 
where one assumes that one of the couplings 
is much smaller than the other and expands 
the phinon frequencies in powers of $k_1/k_2$ or 
$k_2/k_1$, respectively. 
Such a calculation can be performed, for example, in the 
mean field approximation. 

Due to the translational invariance of the crystal, 
the vibrational modes must be characterized 
by an additional parameter: their wave-vector $\vec q$. 
It is defined by the relation 
\begin{eqnarray} 
u_i(l,s) = u_i(\vec q,s) e^{i[\vec q \vec r (l,s)-\omega t]}
\label{eq38u7}
\end{eqnarray}
where $i=1,2,3$ and $\vec r(l,s)$ represents the equilibrium position 
of atom s in primitive cell l. Using this as an ansatz one may solve the 
equations of motion and gets 3n solutions with frequencies 
$\omega(\vec q, \alpha)$, $\alpha=1,...,3n$, where n is the number 
of atoms in the unit cell. For an infinite crystal 
a band structure is obtained for each mode. 

The total vibrational free energy $V_F = \sum_{\alpha} V_\alpha$ 
is a sum of contributions from the modes $\alpha$ which are given by 
\begin{equation} 
V_\alpha = k_B T \sum_{\vec q} \ln (2 \sinh \frac{\hbar \omega(\vec q,\alpha)}{2k_B T})
\label{e668}
\end{equation}
where $\omega(\vec q,\alpha)$ is the frequency of the $\alpha$-th mode 
at wave-vector $\vec q$.
In an infinite crystal one should instead consider the free energy 
per primitive cell, also called $V_\alpha$ here and given by:
\begin{equation} 
V_\alpha = \frac{k_B T V}{(2\pi)^3} \int_{BZ} 
d^3 q \ln (2 \sinh \frac{\hbar \omega(\vec q,\alpha)}{2k_B T})
\label{e669}
\end{equation}
At low temperatures these expressions, which are non-analytic in 
T, have only a very weak temperature dependence. In fact at small T
eq. (\ref{e668}) can be approximated by 
\begin{equation} 
V_\alpha (T\rightarrow 0) = \frac{1}{2} \sum_{\vec q} \hbar \omega(\vec q,\alpha)
+ k_B T \exp [-\frac{\hbar \omega(\vec q,\alpha)}{k_B T}] 
\label{e668aa}
\end{equation}
Note that since I want to identify quarks and leptons with the 
vibrational modes, eq. (\ref{e668aa}) is a direct 
measure of their mass, at least in this idealized situation 
of a translationally invariant internal crystal. 
One may speculate at this point about the definition 
of temperature in the inner symmetry space and 
whether the same Boltzmann constant should be 
used there to transform 'Kelvin' units into energy. 
A related question is how and if at all entropy is 
exchanged between internal and physical space. 
In any case, the momenta of phinons excited at present days energies 
are rather small as compared to the extension of the 
internal Brillouin zone, with the dependence on temperature 
being very weak. 
We are therefore mostly concerned with
the special case of $\vec q=0$, the vibrational 
behavior of the crystal at the center of the 
Brillouin zone, the so-called $\Gamma$-point. 
For the nonlinear process of scattering of two phinons 
discussed in section 4 one should in principle 
consider balanced composite states where one phinon 
has momentum $+\vec q$ and the other one $-\vec q$ (cf. eq. (\ref{eq3001})), 
and this will involve representations of space groups at 
general points $\vec q$. 


A plane wave in our full 3+3+1 dimensional model is a Bloch wave 
\begin{equation} 
\Psi_{ia}= 
f_a(\vec x) u_\alpha(\vec r) 
e^{i(\vec p \vec x - \nu t)} 
e^{i(\vec q \vec r - \omega t)}
\label{exxhg}
\end{equation}
where the amplitude u has the periodicity 
of the internal lattices and the amplitude f has the periodicity 
of the lattice on physical space (provided such a lattice 
exists, c.f. footnote 1). 
It is a bosonic amplitude with respect to internal space 
(vector $\vec r$, index $\alpha$ and 
momentum $\vec q$ for modes $\alpha=1,...,24$) 
and a fermion wave with respect to 
the lattice on physical space (vector $\vec x$, 
index a and momentum $\vec p$) whose spacing however 
is assumed to be too small to be observable experimentally, 
so that Poincare invariance may be safely assumed. 
If one wants to discuss effects from the lattice structure 
in physical space, one should choose the Bloch functions 
in such a way that they transform according to the 
spin-$\frac{1}{2}$ representation of the relevant 
point group of the physical lattice. This 2-dimensional 
representation is usually called $G_1$, and its 
effects within the framework of the tetron model 
have been analyzed in ref. \cite{lampec}.


The spinor index a runs from 1 to 2 or from 1 to 4, depending 
on whether one considers relativistic effects or not. 
For simpler understanding I usually restrict 
myself to the nonrelativistic picture. 
However, by using Dirac fields instead of 
2-component Pauli spinors one may easily 
include antiparticles in the considerations. 


Considering plane waves for the internal vibrations 
may be dangerous in view of the fact that nothing 
is known about the forces acting between lattice points, 
and at later stages strong anharmonicity properties 
will indeed be assumed. 
However, Bloch waves do not have this problem 
because Blochs theorem presupposes only translational 
invariance and holds independently of the nature of the forces.\footnote{Note
again that we are talking about Bloch waves in internal space 
while the Bloch waves in physical space 
reduce to ordinary free fermion fields in the 
limit of large distances (small spatial lattice constants).} 

Nevertheless, it is difficult to estimate the physical significance 
of the internal wave vector $\vec q$ and the associated dispersion. 
In eq. (\ref{e669}) $\vec q$ appears as an integration variable, 
i.e. it is integrated out when it comes to the calculation of 
total energies as measured from physical space. 
Geometrically, $\vec q$ corresponds to a translatory motion of 
the unit cell on the inner symmetry lattice, and one may therefore 
justly ask the question, how extended the inner lattice is, 
i.e. whether it is infinite (with a dependence of the modes on 
a continuous $\vec q$, which an observer on the base space 
cannot measure) or if it consists only of a few unit cells, e.g. 
with periodic boundary conditions. What I have in mind 
is the idea, that through some compactification process  
the crystal consists maybe of only one unit cell, 
in which case there is trivially only 
one wave vector in the first Brillouin zone
and the 'crystal' resembles more an inner molecule.
As discussed earlier, a 3-dimensional lattice with 8 atoms 
in the unit cell has 24 vibrational modes. 
However an 8-atomic molecule has only 18 of them, 
where the reduction by 6 
is due to lacking translational and rotational degrees of freedom. 
A way out in order to keep the 24 modes is to assume that 
the single compact unit cell is physically fixed on its Minkowski 
base point with no large translations or rotations possible. 
In that case the 6 translations and rotations re-establish to become 
vibrational modes again.

The assumption of an internal molecule fixed to its base point 
has several advantages over a large internal crystal. 
Not only one can do without internal wave vectors, 
but one also avoids the problem that in a large internal crystal 
the phinons would dissipate to infinity, 
while in a smaller discrete structure 
they will remain localized and perpetual excitations. 
Furthermore, it is difficult to imagine how a 
large internal crystal is 'fixed' to a base point in 
a way that translational invariance (and the 
validity of Blochs theorem) is not destroyed, 
because internal atoms which are far away from the base point 
may feel much weaker forces than those near to the base. 
A possible way out is that one works with a finite number of 
unit cells which are distributed symmetrically 
with respect to their attractive base point so that one can 
use periodic boundary conditions. In that case 
one has only a finite number of q-values to consider, 
and the sums e.g. in eq. (\ref{e668}) go over those values.  

Note that these considerations also affect the question of 
acoustic modes and massless excitations. Usually, in phonon theory 
the energy of 3 of the 3n phonon branches goes to zero as 
$\vec q \rightarrow 0$. These are called
the acoustic modes and are equivalent to pure translations 
of the crystal. 
According to the last column of table 1 in the present case 
they correspond to one of the $T_g$ representations in eq. (\ref{eq8hg}). 
However, if the finite internal crystal or the one-cell molecule is 
fixed to the base point by some harmonic force the acoustic modes 
will acquire a non-zero energy even at q=0.

The question lattice or single unit cell will be left 
open for the rest of the paper. Although I find the 
one-unit-cell option somewhat simpler to imagine, 
because in that case each real point in physical space 
just splits into 8 vibrating atoms, 
I will often concentrate on the phinon picture with 
more than one unit cell, because it has a richer phenomenology 
and also allows for phase transitions.


\section{Gauge Symmetries from Phinon Scattering}

If one accepts the above ideas about fermions, the immediate question is
how the gauge interactions can arise from a discrete inner symmetry. 

In the present context they are to be interpreted as anharmonic phinon 
scattering states, with the full gauge symmetries appearing as 
emergent phenomena. 
Just as the quarks and leptons themselves the gauge bosons are 
vibrational excitations which however arise 
from nonlinear forces between the atoms 
of the inner symmetry lattice, corresponding to 
anharmonic terms typically of fourth order in the phinon interactions 
which in turn allow for scattering processes among the phinons. 
Thus the emergence of gauge structures consists in two steps: firstly, 
the gauge bosons arise as bound states from the 
scattering processes of phinons, and secondly 
their effective interactions should follow a gauge symmetry, 
in order to keep track of the 'connection' and the 
differential geometry of the full 3+3+1 dimensional 'fibre space'. 

A similar point of view has been taken in ref. \cite{lampe1} 
where it was shown that on a group theoretical level gauge bosons can be 
constructed as 'Cooper states' of the quark and lepton 
excitations. The idea is in fact not unusual 
and is used both in relativistic particle physics\cite{hase,dimo,yana,amat}
and in non-relativistic statistical models\cite{mara,seme,berr,cord} 
with continuous as well as with discrete base spaces. 
In the present framework the starting point would be 
the nonlinear phinon and mignon interactions eqs. (\ref{eq331g}) 
and (\ref{em15}) given below 
which lead to scattering processes 
in the internal crystal. On the base space this will induce 
an effective Lagrangian
of the generic form 
\begin{equation} 
L=\sum_{n,m} \bar f (\vec n) \gamma^\mu G_\mu^a (\vec n,\vec m) \lambda^a f(\vec m)
\label{e66690}
\end{equation}
where $\gamma_\mu$ are the ordinary Dirac matrices in 3+1 dimensions, 
$\lambda^a$ the generators of the local gauge group 
and the sum is over (base space) lattice sites n and m. 
The interacting field is constructed as 
\begin{equation} 
G_\mu^a(\vec n)=<0|\bar f (\vec n) \gamma_\mu \lambda^a f(\vec m + \vec \mu)|0> 
\label{e66190}
\end{equation}
(with $\vec \mu$ being the next lattice point in the $\mu$-direction) 
and behaves as 
\begin{equation} 
G_\mu(\vec n)\rightarrow U(\vec n)G_\mu(\vec n)U^\dagger( \vec m + \vec \mu)
\label{e62690}
\end{equation}
under local gauge transformations while fermions transform as 
\begin{equation} 
f (\vec n)\rightarrow U(\vec n) f (\vec n) 
\label{e61690}
\end{equation} 
It can then easily be proven that the Lagrangian eq. (\ref{e66690}) 
is gauge invariant. 

The method has also been used \cite{berr,seme,mara} 
to prove the equivalence of Heisenberg spin models with certain
gauge theories and can indeed be generalized: if the scattering states 
exist, then gauge groups with a continuous symmetry appear as necessary 
by-products of the model. 
For the case of the tetron model 
these arguments have been presented in detail in ref. \cite{lampe1}, 
and I do not want to repeat them here for the 
complete Standard Model, but just sketch the prove for the simpler 
case of QCD interactions: from eq. (\ref{eq8hg}) one sees that 
quarks always appear as triplets. 
One would like to identify these triplets with SU(3) color triplets. 
In order that to be possible one has to assume that gauge 
symmetries and interactions are emergent effects 
which arise from the originally discrete symmetry. 
In a first step one may analyze the symmetry content 
of a $2\rightarrow 2$ scattering of $A_4$-triplets: 
\begin{equation} 
T \otimes T = A+A'+A''+2T
\label{e6694}
\end{equation}
Using the symmetry adapted functions for the representations 
T on the left hand side of (\ref{e6694}) and evaluating the corresponding 
Clebsch-Gordon coefficients leads to 
\begin{itemize}
\item
a representation of the $(B-L)$-photon as 
\begin{eqnarray}  
B_\mu=\bar u_1 \gamma_\mu u_1+\bar u_2 \gamma_\mu u_2
+\bar u_3 \gamma_\mu u_3 
\label{eqpho5}
\end{eqnarray} 
which originates from the representation 
$A$ on the right hand side of 
eq. (\ref{e6694}) and from the corresponding 
Clebsch-Gordon coefficient \cite{griff}
\begin{eqnarray}  
V(T,T,A;i,j,1)=\frac{1}{\sqrt{3}} \delta_{ij} 
\label{ewqpho5}
\end{eqnarray} 
\item 
a representation of the gluon octet stemming from 
the remaining part $A'+A''+2T$ of 
the decomposition eq. (\ref{e6694}). Namely, 
the CG-coefficients can be written in terms of 
the Gell-Man $\lambda$-matrices as 
\begin{eqnarray}  
V(T,T,A';i,j,1)&=&\frac{1}{2}\lambda_{8ij}\\
V(T,T,A'';i,j,1)&=&\frac{1}{2}\lambda_{3ij}\\
V(T,T,T;i,j,k)&=&\frac{1}{\sqrt{6}} \epsilon_{ijk} \\
                    &=&\frac{i}{\sqrt{6}}\lambda_{7,5,2 ij} \quad for \quad k=1,2,3
\label{eqgg145}
\end{eqnarray} 
where the last expression corresponds to the first copy 
of T on the right hand side of eq. (\ref{e6694}) and 
\begin{eqnarray}  
V(T,T,T;i,j,k)&=&\frac{1}{\sqrt{6}} |\epsilon_{ijk}|\\
                    &=&\frac{1}{\sqrt{6}}\lambda_{6,4,1 ij} \quad for \quad k=1,2,3
\label{eqgg148a}
\end{eqnarray} 
to the second.
All in all we obtain 
\begin{eqnarray}  
G_{3\mu}&=&\bar q_1 \gamma_\mu q_1-\bar q_2 \gamma_\mu q_2 \\
G_{8\mu}&=&\frac{1}{\sqrt{3}}(\bar q_1 \gamma_\mu q_1
    +\bar q_2 \gamma_\mu q_2-2\bar q_3 \gamma_\mu q_3) 
\label{eqgg45}
\end{eqnarray} 
and similarly for the other $\lambda$-matrices. 

The fact that formally the same bilinear combinations are created 
as needed in $SU(3)_{color}$-QCD is no accident but has 
to do with the fact that $S_4$ and $A_4$ may be considered as 
subgroups of SO(3) $\subset$ SU(3). 
The result is therefore an 
elaboration on the claim formulated in \cite{lampe2} that the 
appearant tetrahedral symmetry of quarks and leptons is able 
to provide 'germs' of the Standard Model interactions, 
which are necessarily gauge interactions because these are the only 
consistent quantum field theories fulfilling the appropriate 
symmetry requirements (at energies small as compared to 
the inverse lattice spacings).  
\end{itemize}

Similarly the weak bosons can be obtained from the 
$Z_2$ factor of the pyritohedral group as 
$g\otimes g$, $g\otimes u - u\otimes g$ and $u\otimes u$, 
with an additional contribution $g\otimes u + u\otimes g$ 
to the $(B-L)$-photon. 
A tricky point, however, is to understand parity violation of the weak 
interactions, i.e. the fact that only lefthanded fermions 
take part in the corresponding processes. 
This will be treated in section 6, where a relation 
between the chirality of the internal lattice 
and weak parity violation will be constructed.

Another complication is the possible presence of an internal 
wave vector. If the internal crystal consists only of one unit cell, 
the above analysis is complete and there is nothing more to be said. 
However, if it is an extended crystal with 
wave vectors q, the phinon-phinon interactions, which lead 
to the gauge fields, will involve scattering 
processes $(\vec q)\otimes (-\vec q)$. 
This will be made concrete in the next section where 
the nonlinear terms responsible for the scattering will 
be written down explicitly (cf. eq. (\ref{eq331g})).
As a consequence, the group theoretical analysis presented above 
will become more 
complicated, because one has to consider representations 
of the full space group and not just of the point group $A_4\times Z_2$. 
These depend on the value of $\vec q$, on its star and the 
associated little groups, and the corresponding tensor products 
must be looked up, for example, on the Bilbao 
crystallographic server \cite{bilbao,bilbao1}. 


\section{Chirality and the Breakdown of the Tetrahedral Symmetry}

In the Standard Model the quark and lepton masses together with the 
Kobayashi-Maskawa mixing elements are free parameters, while 
they should be calculable in the tetron model as 
vibrational energies of phinon modes. 

Experimentally there is a large difference in the masses of fermions, 
starting from (almost) massless neutrinos to the top quark mass. 
There are large differences between the families, but also 
within one family, and most noticeable even 
among weak isospin partners. 
The corresponding breaking of weak SU(2) does not 
mean that pyritohedral $A_4 \times Z_2$ is 
a broken symmetry, too. On the contrary, it is unbroken, because 
it was constructed in such 
a way that all different mass particles can be ordered 
in different mass multiplets of this point group 
(cf. eq. (\ref{eq8hg})).

A natural question to ask is how a rather 
ugly symmetry like $A_4 \times Z_2$ may have arisen 
to become the fundamental ordering principle of nature. 
After all, $A_4\times Z_2$ (and also $A_4$) is a chiral point group, 
and a crystal with such a symmetry always shows a chirality 
structure in the form of a (left or right handed) helical 
arrangement of the crystallic atoms.
Therefore it seems reasonable to assume that $A_4 \times Z_2$ 
itself arose from a primordial symmetry breaking within the inner 
symmetry crystal, i.e. that soon after the big bang there 
was a second order phase transition from a high-temperature phase 
with a larger lattice symmetry (tetrahedral $S_4$ or even the complete 
octahedral group $O_h$) to the present $A_4 \times Z_2$ or $A_4$ 
phase. Correspondingly, the original multiplets would break up as 
\begin{eqnarray} 
A_{2g}+A_{2u}+E_u+E_g+2T_{1g}+2T_{1u}+T_{2g}+T_{2u} \nonumber \\  
\rightarrow A_g+A'_g+A''_g+3T_g + A_u+A'_u+A''_u+3T_u
\label{eq33hg}
\end{eqnarray}
for the breaking $O_h \rightarrow A_4\times Z_2$ leading to the 
desired fermion spectrum eq. (\ref{eq8hg}) or according to  
\begin{equation} 
A_1+A_2+2E+3T_1+3T_2 \rightarrow 2(A+A'+A''+3T)
\label{eq3phg}
\end{equation}
for $S_4 \rightarrow A_4$.\footnote{An alternative 
to spontaneous symmetry breaking 
is an {\it explicit} breaking of symmetries, 
which could be induced for example by a pseudoscalar 
chiral interaction among the lattice atoms. 
Such an interaction is given e.g. by the scalar triple product 
\begin{equation} 
H_{3}= f_3 \sum_{l,l',l'',s,s',s''} \vec u (l,s) [\vec u(l',s') 
\times \vec u(l'',s'')] 
\label{eq3334}
\end{equation}
of lattice vectors which is positive for even permutations 
and negative for odd ones. In the philosophy discussed 
in the main text, however, eq. (\ref{eq3334}) is merely 
an effective interaction which could arise after 
the spontaneous symmetry breaking and may be used 
to describe chiral effects in the low energy regime.}  
Note that the left hand side 
of eq. (\ref{eq3phg}) corresponds to expression (\ref{eq7hg})
while the left hand side of eq. (\ref{eq33hg}) 
is obtained in a similay manner for a crystal with point group $O_h$ 
and 8 atoms in the unit cell.

How do these symmetry breakings come about? 
Since we have identified phinons as the 
relevant excitations it is tempting to suppose 
that we have to look for a {\it displacive} phase 
transition on the internal symmetry lattice, i.e. 
a spontaneous effect where the crystal is 
deformed by one phinon 'freezing' out at 
a critical temerature $T_c$\cite{coch,sal,soll}. 
For this to happen, one needs a strongly anharmonic 
interaction, because the phinon frequency must fall to zero 
at the critical point, and then 
harden again below it, as the crystal finds a new 
equilibrium around the deformed structure. 

Note that in the present case such a transition, which  
may or may not be accompanied by a 
compactification of the lattice (as discussed 
in the previous section), would be 
from a non-chiral to a chiral inner symmetry lattice. 
In the following this process will be described 
in some detail: consider the universe shortly after the big bang 
where the temperatures were high and the world was in 
a phase of high symmetry with all sorts of phinons 
excited. When the temperature fell, the 
frequency of one (or some) of the modes decreased,  
and when its value within the 
inner symmetry crystals fell below the critical 
point, this mode froze out to zero frequency and 
transformed to a static displacement pattern, 
thus inducing a transition to $A_4\times Z_2$ or $A_4$ 
and a shift in the Wyckoff positions to their present values.

Displacive phase transitions are well known in 
solid state physics and are usually described 
in terms of the normal coordinates $u$ used to describe the 
soft vibrational modes. In other words, 
the order parameter to be chosen 
is the displacement vector of the mode 
which freezes out at the phase transition, 
or, more generally, the mean displacement 
\begin{equation} 
\eta= \frac{1}{N} \sum \vec u(l,s)
\label{eqmi694}
\end{equation} 
A simpler definition can be obtained 
by comparing the fractional coordinates of the atoms in
the low-temperature and high-temperature phases. 
This has the further advantage that one circumvents 
all problems associated with thermal expansion 
which arise when using absolute displacements. 
For example, if atoms with fractional coordinates x, ... 
in the low-temperature phase tend towards $\frac{1}{4}$, ... 
on heating to the high-temperature phase a natural 
order parameter would be 
\begin{equation} 
\eta=\frac{1}{4}-x 
\label{eqaiaa}
\end{equation} 
If, on the other hand, a transition
involves rotations of a group of atoms, 
the order parameter can just as well be defined 
as the angle by which the group of atoms has rotated to break the symmetry. 
In leading order and for small displacements all 
these definitions will be equivalent. 
However, some of the displacements
and rotations may follow the order parameter to higher
order. Therefore it is better not to 
simply average over anything as in eq. (\ref{eqmi694}),
but to select those which are important in 
the formation of the new structure. 

In the case at hand the freeze out proceeds 
in such a way that the achiral reflection symmetries 
gets lost: 
\begin{itemize}
\item For $S_4 \rightarrow A_4$: the symmetric group 
$A_4$ is the group of proper rotations of a equilateral 
tetrahedron, while the tetrahedral group $S_4$ contains 
rotoreflections in addition. If $S_4$ is interpreted 
as the group of permutations of 4 objects, 
these rotoreflections correspond to the odd permutations 
in $S_4$, while $A_4$ contains even permutations only. 
The possible phase transitions for crystals with 8 atoms 
in the unit cell are: 
218 (P43n) $\rightarrow$ 195 (P23), 219 (F43c) $\rightarrow$ 196 (F23) 
and 220 (I43d) $\rightarrow$ 199 (I213)\cite{bilbao1}.
For example, in case 218 according to eq. (\ref{eq3uijj}) 
the symmetric high temperature phase is characterized by 
2 tetrahedrons in the unit cell, both of extension x, 
which are shifted uniformly 
by a vector $(\frac{1}{2},\frac{1}{2},\frac{1}{2})$ 
whereas the broken phase is characterized by 
a transition of Wyckoff positions $8e \rightarrow 4e_1+4e_2$ with $x_1=x$ 
and $x_2=x+\frac{1}{2}$, i.e. one is forced to introduce 
2 types of Wycks in order to obtain the 2 required 
sets of representations $A+A'+A''+3T$.
There is a similar situation for space groups 219 and 220 
where one finds $32e \rightarrow 16e_1 + 16e_2$ 
and $16c \rightarrow 8a_1 + 8a_2$, respectively 
(cf. eq. (\ref{eq3zw44})).
\item For $O_h\rightarrow A_4\times Z_2$ (or $O_h\rightarrow A_4$): the 
octahedral group $O_h = O\times P_{in}$ is in fact the direct 
product of proper rotations of a cube and the (inner) parity 
operation $P_{in}$, so that the breaking to $A_4\times Z_2$ 
includes a breaking of $P_{in}$. 
This breaking has the further advantage, that for a 
proper choice of space groups one can come along with 
only one type of Wyckoff positions. 
I have systematically scanned all space groups 
and Wyckoff positions in question and found only two 
solutions which fulfill all requirements: 
\begin{itemize}
\item Firstly there is 
\begin{equation} 
32b(228=Fd3c) \rightarrow A_{2g} \rightarrow 32e(203=Fd3) 
\label{eq228}
\end{equation} 
where the representation between the arrows denotes the 
mode that freezes out at the phase transition. 
For the starting configuration 228=Fd3c (32b) the Wyckoff positions are
\begin{eqnarray} 
F_1=(\frac{1}{4},\frac{1}{4},\frac{1}{4}) \quad & & F_2=(0,\frac{1}{2},\frac{3}{4}) \nonumber \\ 
F_3=(\frac{1}{2},\frac{3}{4},0) \quad & & F_4=(\frac{3}{4},0,\frac{1}{2}) \nonumber \\ 
F_5=(\frac{3}{4},\frac{3}{4},\frac{3}{4}) \quad & & F_6=(0,\frac{1}{2},\frac{1}{4}) \nonumber \\ 
F_7=(\frac{1}{2},\frac{1}{4},0) \quad & & F_8=(\frac{1}{4},0,\frac{1}{2}) 
\label{ew22832b}
\end{eqnarray}
On cooling below the critical temperature these will go over to the positions
\begin{eqnarray} 
G_1=(x,x,x) \quad & & G_2=(-x+\frac{1}{4},-x+\frac{1}{4},x) \nonumber \\ 
G_3=(-x+\frac{1}{4},x,-x+\frac{1}{4}) \quad & & G_4=(x,-x+\frac{1}{4},-x+\frac{1}{4}) \nonumber \\ 
G_5=(-x,-x,-x) \quad & & G_6=(x+\frac{3}{4},x,x+\frac{3}{4},-x) \nonumber \\ 
G_7=(x+\frac{3}{4},-x,x+\frac{3}{4}) \quad & & G_8=(-x,x+\frac{3}{4},x+\frac{3}{4}) 
\label{ew20332e}
\end{eqnarray}
of Fd3=203 (32e) with point group $A_4\times Z_2$. The representations 
describing the phinon modes transform as given in eq. (\ref{eq33hg}) and 
the order parameter may be chosen as anticipated in eq. (\ref{eqaiaa}). 
\item Secondly there is 
\begin{equation} 
16b(230=Ia3d) \rightarrow A_{2g} \rightarrow 16c(206=Ia3)
\label{eq23000}
\end{equation} 
In this case the higher symmetry phase 230=Ia3d (16b) has the configuration 
\begin{eqnarray} 
D_1=(\frac{1}{8},\frac{1}{8},\frac{1}{8}) \quad & & D_2=(\frac{3}{8},\frac{7}{8},\frac{5}{8}) \nonumber \\ 
D_3=(\frac{7}{8},\frac{5}{8},\frac{3}{8}) \quad & & D_4=(\frac{5}{8},\frac{3}{8},\frac{7}{8}) \nonumber \\ 
D_5=(\frac{1}{8},\frac{1}{8},\frac{1}{8}) \quad & & D_6=(\frac{3}{8},\frac{7}{8},\frac{5}{8}) \nonumber \\ 
D_7=(\frac{7}{8},\frac{5}{8},\frac{3}{8}) \quad & & D_8=(\frac{5}{8},\frac{3}{8},\frac{7}{8})  
\label{ew23016b}
\end{eqnarray}
which on cooling below the critical temperature go over to the positions
\begin{eqnarray} 
E_1=(x,x,x) \quad & & E_2=(-x+\frac{1}{2},-x,x+\frac{1}{2}) \nonumber \\ 
E_3=(-x,x+\frac{1}{2},-x+\frac{1}{2}) \quad & & E_4=(x+\frac{1}{2},-x+\frac{1}{2},-x) \nonumber \\ 
E_5=(-x,-x,-x) \quad & & E_6=(x+\frac{1}{2},x,-x+\frac{1}{2}) \nonumber \\ 
E_7=(x,-x+\frac{1}{2},x+\frac{1}{2}) \quad & & E_8=(-x+\frac{1}{2},x+\frac{1}{2},x)  
\label{ew20616c}
\end{eqnarray}
so that the order parameter may be chosen to be 
\begin{equation} 
\eta=\frac{1}{8}-x 
\label{eqaiia}
\end{equation} 
\end{itemize}
\end{itemize}

The phase transitions described above will remove 
all degenericies between isospin partners, 
as contained in the representations E on 
the left hand sides of (\ref{eq33hg}) and (\ref{eq3phg}). 
Since the weak vector bosons are excitations between 
isospin partners and therefore between 
odd and even permutation states, the symmetry 
breaking will give masses to W and Z boson as well
and thus will eventually be responsible for the $SU(2)_L$ 
breaking which in the Standard Model is described 
by the Higgs mechanism. 


Using the displacement of the freeze out mode
as order parameter $\eta$ one may apply the Landau theory of phase 
transitions to obtain the symmetry breaking part of the free energy 
which in general reads
\begin{equation} 
V(\eta)= -\frac{1}{2} f_2 \eta ^2 + \frac{1}{4} f_4 \eta^4 +O(\eta^6)
\label{eqii694}
\end{equation}
with a positive parameter $f_4$ 
and $f_2 =a (T-T_c)$ being linear in $T-T_c$ and positive 
for T larger than $T_c$. 
Odd powers of $\eta$ are excluded from eq. (\ref{eqii694}) by symmetry 
arguments and higher order terms can usually be neglected. 
The magnitude of the distortion in this potential is given by 
\begin{equation} 
|\eta_0|= \sqrt{\frac{f_2}{f_4}}
\label{eqmm694}
\end{equation}

On the microscopic level the Landau free energy eq. (\ref{eqii694}) 
can be interpreted as the local double-well potential in which 
the inner atoms move and which is the 
driving force behind the phase transition. 
Originally it comes from anharmonic terms in the 
Lagrangian and is responsible for the freeze out of the 
soft mode. For a proper treatment 
the analysis should be extended to include all modes, 
i.e. all displacement coordinates. 
For simplicity I will assume that a transformation 
to normal coordinates has been performed and that these 
are given in momentum space as $u(\vec q,\alpha)$, 
$\alpha=1,...,3n$ with frequencies $\omega(\vec q,\alpha)$.
If anharmonic interactions for all modes are included to fourth order, the 
interaction part of the Hamiltonian is given by 
\begin{eqnarray} 
 H_W&=&\frac{1}{2} \sum_{\vec q,\alpha} \omega^2(\vec q,\alpha) u(\vec q,\alpha) u(-\vec q,\alpha) 
   +\frac{1}{4!} \sum_{\vec q_1,\vec q_2,\vec q_3,\vec q_4,\alpha} 
    \lambda (\vec q_1,\vec q_2,\vec q_3,\vec q_4,\alpha) 
              \nonumber \\  & &      u(\vec q_1,\alpha) u(\vec q_2,\alpha) 
                                    u(\vec q_3,\alpha) u(\vec q_4,\alpha)
                  \delta (\vec q_1+\vec q_2+\vec q_3+\vec q_4)
\label{eq331g}
\end{eqnarray}
In the approximation of the mean field theory 
part of the normal mode coordinates may be  
replaced by their thermal average to obtain an effective interaction:
\begin{eqnarray} 
H_W&=&\frac{1}{2} \sum_{\vec q,\alpha} \omega^2(\vec q,\alpha) u(\vec q,\alpha) u(-\vec q,\alpha) \nonumber \\ 
   &+&\frac{1}{4} \sum_{\vec q,\vec q',\alpha} \lambda (\vec q, -\vec q, \vec q', -\vec q',\alpha) 
                <u(\vec q',\alpha) u(-\vec q',\alpha)> u(\vec q,\alpha) u(-\vec q,\alpha)
\label{eq33332g}
\end{eqnarray}
Here the mean fields $<u(\vec q_1,\alpha) u(\vec q_2,\alpha)>$ 
vanish due to crystal momentum conservation 
except for $\vec q_1 = -\vec q_2$. 

It should be noted that the terms 
$<u(\vec q',\alpha) u(-\vec q',\alpha)> u(\vec q,\alpha) u(-\vec q,\alpha)$ 
give the internal contributions to the required 
interaction Hamiltonian eq. (\ref{e66690}) on the base space 
and thus to the desired gauge theories. 
Furthermore, these terms can be used to describe the phase transition 
induced by the $\eta$-mode: 
since eq. (\ref{eq33332g}) is quadratic in the normal coordinates 
one may define effective frequencies $\Omega$ by 
\begin{eqnarray} 
\Omega^2(\vec q,\alpha)= \omega^2(\vec q,\alpha) 
    +\frac{1}{2} \sum_{\vec q'} \lambda (\vec q, -\vec q, \vec q', -\vec q',\alpha) 
                <u(\vec q',\alpha) u(-\vec q',\alpha)> 
\label{eq3001}
\end{eqnarray}
which fulfill the quasi-harmonic relation 
\begin{eqnarray} 
H_W= \frac{1}{2} \sum_{\vec q,\alpha} \Omega^2(\vec q,\alpha) u(\vec q,\alpha) u(-\vec q,\alpha)
\label{eq334g}
\end{eqnarray}
The redefined frequencies contain the leading order anharmonic effects 
and also the temperature dependence eq. (\ref{eq17744}) 
which will enter the description of the phase transition. 
The point is that the mean field $<u(\vec q,\alpha) u(\vec q',\alpha)>$ 
is the Fourier transform of the 
correlation function\cite{dove2} and therefore can be given in terms of the 
phinon occupation number  
\begin{eqnarray} 
n(\Omega(\vec q,\alpha),T)= \frac{1} {\exp(\frac{\hbar \Omega(\vec q,\alpha)}{k_B T}) + 1}
\label{em2}
\end{eqnarray}
as  
\begin{eqnarray} 
<u(\vec q,\alpha) u(\vec q',\alpha)>
=\frac{\hbar}{\Omega(\vec q,\alpha)} [n(\Omega(\vec q,\alpha),T)+\frac{1}{2}]
\label{em1}
\end{eqnarray}
For $k_B T \geq \hbar \Omega(\vec q,\alpha)$ this is given to 
a very good approximation as 
\begin{equation} 
<u(\vec q',\alpha) u(-\vec q',\alpha)> = \frac{k_B T}{\Omega^2(\vec q',\alpha)}
\label{eq17744}
\end{equation} 

To get a physical understanding let me assume 
that only one of the normal modes is 
responsible for the deformation and call the corresponding 
normal coordinate $\eta$. 
In fact this corresponds to the $A_{2g}$ mode 
in the cases under consideration. In the high temperature phase 
it is just like any other normal mode, whereas near the 
critical point it freezes out and may be taken as 
the order parameter. Accordingly, eq. (\ref{eq33332g}) 
can be put in the form
\begin{eqnarray} 
H_W=V(\eta)+ \frac{1}{2} \sum_{\vec q,\alpha} [\omega^2(\vec q,\alpha) 
+\frac{1}{2} g_4(\vec q) \eta^2 ] 
                    u(\vec q,\alpha) u(-\vec q,\alpha) 
\label{eq336g}
\end{eqnarray}
where the quartic coupling of the soft mode to the 
other modes has been abbreviated as $g_4$.

The first term $V(\eta)$ in this equation contains 
the self interactions of the frozen mode. 
The corresponding couplings were called $-f_2=a (T_c-T)$ and $f_4$ 
in eq. (\ref{eqii694}) 
instead of $\omega^2$ and $\lambda$, and one sees that the 
frequency becomes imaginary at the transition point. 
The physical reason is that at low temperatures 
the structure of the high temperature phase is unstable 
with respect to the distortion of the soft mode 
and therefore the harmonic term of the 
mode must have a maximum for $\eta=0$.
This can happen only, if the phinon frequency is imaginary. 
Above the transition point the 
high-symmetry phase is stable, because the 
anharmonic interactions increase the 
phinon frequency on increasing the temperature 
and the soft mode is an ordinary stable normal mode there.

To make contact between the microscopic theory and the 
Landau approach to phase transitions one can 
expand the standard expression for the 
phinon free energy eq. (\ref{e668}) in powers of $\eta$: 
\begin{eqnarray} 
V_F=V_F(\eta=0)+\frac{1}{4} k_B T \sum_{\vec q} 
    \frac{g_4(\vec q)}{\Omega^2(\vec q)}
    \eta^2 +O(\eta^4)
\label{eq339g}
\end{eqnarray}

The 'Landau free energy' is defined as the contribution of the order parameter 
to the free energy. 
Taking the $\eta^2$ term in (\ref{eq339g}) as correction to the 
potential $V(\eta)$ eq. (\ref{eqii694})
one obtains the result for the 
Landau free energy:  
\begin{eqnarray} 
V_L(\eta)= -\frac{1}{2} f_2 \eta^2 
+  \frac{1}{4} k_B T \sum_{\vec q} 
\frac{g_4(\vec q)}{\Omega^2(\vec q)} T \eta^2
+ \frac{1}{4} f_4 \eta^4 +O(\eta^6)
\label{eq349g}
\end{eqnarray}
which can easily be put into the standard form 
\begin{eqnarray} 
V_L(\eta)= -\frac{1}{2} a(T-T_c) \eta^2 
+ \frac{1}{4} f_4 \eta^4 +O(\eta^6)
\label{eq349g61}
\end{eqnarray}
where
\begin{eqnarray} 
a=\frac{1}{4} k_B \sum_{\vec q} 
    \frac{g_4(\vec q)}{\Omega^2(\vec q)}
\label{eq44349g}
\end{eqnarray}
and $T_c=-\frac{f_2}{a}$.

The model also can be used to calculate the behavior of 
the Landau free energy at low temperatures. 
Namely, the complete free energy is the sum of eqs. 
(\ref{eqii694}) and (\ref{e668})
\begin{equation} 
V_C = -\frac{1}{2} f_2 \eta^2 + \frac{1}{4} f_4 \eta^4
+k_B T \sum_{\vec q,\alpha} \ln (2 \sinh \frac{\hbar \omega(\vec q,\alpha)}{2k_B T})
\label{e669ee}
\end{equation}
which in the low-T limit takes the form:
\begin{equation} 
V_C = -\frac{1}{2} f_2 \eta^2 + \frac{1}{4} f_4 \eta^4
+\frac{1}{2} \sum_{\vec q,\alpha} \hbar \Omega(\vec q,\alpha)
-k_B T \sum_{\vec q} \exp [-\frac{\hbar \Omega(\vec q)}{2k_B T}]
\label{e671ee}
\end{equation}
This function takes its minimum at 
\begin{equation} 
\eta^2 = \frac{f_2}{f_4} 
      - \frac{\hbar}{4f_4} \sum_{\vec q} \frac{g_4(\vec q)}{\Omega(\vec q)}
- \frac{\hbar}{4f_4} \sum_{\vec q} \frac{g_4(\vec q)}{\Omega(\vec q)} 
         \exp [-\frac{\hbar \bar\omega(\vec q)}{2k_B T}]
\label{e671ge}
\end{equation}
where the first term is simply the minimum of the original double well potential, 
the second term corresponds to quantum corrections of the zero-point phinon motion 
at $T=0$ and the third term represents the leading temperature 
dependance at low temperatures. 
The quantum correction (second term) lowers the minimum, 
and increasing the temperature it is lowered even further 
up to $T=T_c$ where $\eta^2=0$.

I finish this section with the remark that
the equations above hold only for the case that all 
entropy of the system is vibrational entropy. 
If there is a configurational order-disorder 
contribution to the phase transition\cite{soll,sal} 
this adds another term $\sim \eta^2$ to the 
Landau free energy (very similar in form to what the phinon 
contributes in the second term to eq. (\ref{eq349g})) 
and will thus modify the value of the transition 
temperature. Order-disorder phase transitions do not 
usually play an important role in vibrational problems 
but are more important in magnetic systems, with 
spin waves as quasi-particle excitations, 
and will thus influence the discussion 
presented in section 7.


\section{Inner and Outer Parity Violation}

Parity violation of the weak interactions was discovered 
more than 50 years ago and still is one of the most 
puzzling effects known in physics. 
The fact that only lefthanded 
fermions (and righthanded antifermions) take part 
in the weak interactions should be considered 
a unique hint that inner symmetry and spacetime 
symmetry must be somehow related or mixed, in such a way 
that there is a parity violating contribution to the cross section 
of the form $\vec p \vec s$ 
where $\vec p$ is a fermion momentum and $\vec s$ its spin vector. 

$\vec p \vec s$ is a pseudoscalar quantity built as a product of a 
vector and an axial vector, where the axial vector(=angular momentum) is itself 
a cross product of two vectors, so that $\vec p \vec s$ has in 
fact the form of a scalar triple product like in eq. (\ref{eq3334}). 
More in general any pseudoscalar contribution to a 
cross section signals parity violation, because 
a pseudoscalar changes sign unter parity\cite{pick}. 
In low energy physics, parity violating effects usually arise 
from helical structures in molecules or crystals and can be described 
by pseudoscalars (similar to the above triple product) 
which measure the degree of parity violation\cite{low1,pick}.

For the case of weak parity violation a microscopic 
interpretation is not available until today but 
there is a very successful {\it effective} description 
in terms of V-A currents 
\begin{eqnarray} 
j^{V-A}_\mu = \bar f_L \gamma_\mu f_L 
      =  \bar f \gamma_\mu \frac{1}{2} (1- \gamma_5) f
\label{eq21g}
\end{eqnarray}
which are built into the Standard Model by the use of  
'chiral fermions' $f_L=\frac{1}{2}(1-\gamma_5) f$ in the Lagrangian 
which exactly correspond to the V-A currents given above. 
Parity violation is thus described by 
explicit symmetry breaking terms, 
while in the left-right symmetric model mentioned earlier 
there is spontaneous breaking of parity via a rather complicated 
Higgs sector\cite{su2su2}. 

As a characteristic example of weak parity violation 
in the Standard Model consider the production and 
decay of a W-boson 
\begin{eqnarray} 
u(p_u)+\bar d(p_d) \rightarrow W^+ \rightarrow \mu^+ (p_l) +\nu_\mu (p_n)
\label{eq2279g}
\end{eqnarray}
We shall work in the rest frame of the W 
and assume that the incoming up-quark beam defines 
the positive x-direction. 
Assuming further relativistic fermions and a scattering angle $\beta$ 
one can parametrize the momenta as 
\begin{eqnarray} 
p_u=\frac{m_W}{2}(1,1,0,0) & & p_d=\frac{m_W}{2}(1,-1,0,0) \nonumber \\
p_n=\frac{m_W}{2}(1,\cos \beta,\sin\beta,0) & & p_l=\frac{m_W}{2}(1,-\cos \beta,-\sin\beta,0) 
\label{eq2880g}
\end{eqnarray}
where a righthanded coordinate system has been introduced with 
unit vectors $\vec e_1$, $\vec e_2$ and $\vec e_3$.

Due to the structure of the V-A currents which corresponds 
to the S-matrix element
\begin{eqnarray} 
<W^+(p_W,\epsilon) | S | u(p_u) \bar d(p_d)> \sim
\bar v(p_d) \epsilon_\mu \gamma^\mu \frac{1}{2}(1-\gamma_5) u(p_u)
\label{eq2230g}
\end{eqnarray}
the $W^+$ is produced with helicity $-1$ 
i.e. its angular momentum is antiparallel to the direction 
$\vec p_u$ of the up-quark and can therefore be described by the 
polarization vector for lefthanded circular polarization 
\begin{eqnarray} 
\vec \epsilon_L =\frac{1}{\sqrt{2}} (0,1,-i)=\frac{1}{\sqrt{2}}( \vec e_2-i\vec e_3)
\label{eq2280g}
\end{eqnarray}

On the other hand, the matrix element for the decay of a W-boson with 
polarization $\vec \epsilon$ into two massless leptons 
\begin{eqnarray} 
<\mu^+(p_l) \nu(p_n)| S |W^+(p_W,\vec\epsilon)> \sim 
\bar u(p_n) \epsilon_\mu \gamma^\mu \frac{1}{2}(1-\gamma_5) v(p_l)
\label{eq22g}
\end{eqnarray}
leads to the rate
\begin{eqnarray} 
\frac{d\Gamma}{d \Omega} (W^+ \rightarrow \mu^+ +\nu) 
=\frac{\alpha m_W}{32\pi \sin^2(\theta_W)}
[1-(\hat p_n \vec\epsilon) (\hat p_n \vec\epsilon^\ast)
               - i(\vec\epsilon\times \vec\epsilon^\ast) \hat p_n]
\label{eq22gh1}
\end{eqnarray}
where $\hat p_n=\vec p_n/|\vec p_n|$ and the parity violating term is 
$-i(\vec\epsilon \times \vec\epsilon^\ast)\vec p_n$ 
and built from the triple product of 3 vectors. 
Note that in the case at hand 
the vector $i(\vec\epsilon_L \times \vec\epsilon_L^\ast)=-\vec e_1$ 
points into the negative x-direction (the direction 
opposite to the direction of motion of the original up-quark). 
Note further, that the product $\vec{p} \vec{s}$ mentioned above 
will appear in the bremsstrahlung process $l^\pm \rightarrow \nu_l W^\pm$ 
if the polarization vector $\vec s$ of the incoming lepton is known and 
that in fact the fermion polarisation vector 
$\vec s$ takes over the role of the cross product 
$i(\vec\epsilon \times \vec\epsilon^\ast)$ in eq. (\ref{eq22gh1}).

Although these approaches and the resulting 
phenomenological predictions are able to describe 
accurately all the measured effects I do not think that 
they give an exhaustive and satisfactory explanation 
of the physics behind the phenomena. 
I will now consider the question 
whether the tetron model can lead us to a deeper understanding. 
What I want to do is to tie weak parity violation to the appearant 
chirality of the inner symmetry crystal. As repeatedly mentioned, 
its point group $A_4\times Z_2$ 
is 'chiral', i.e. it violates the parity of the internal 
space, and I will in fact try to construct in this section a connection 
between this inner chirality and the observed parity violation 
of the weak interaction. 

The chirality of the crystal immediately implies that all phinon modes 
have to respect the helical structure of the system. 
Since after the symmetry breaking the $Z_2$-factor 
has nothing to do with inner parity any more, 
both the g- and the u-states in eq. (\ref{eq8hg}) 
will reflect the inner helical structure. 
But how can this be related to 
external weak parity violation which 
is measured on chiral particles in physical space?

In order to answer this question one has to keep in mind, 
that internal and Minkowski space 
originate from a unified (6+1)-dimensional spacetime, 
which at the big bang has split into a physical and 
an inner symmetry part. 
Afterwards, there was the symmetry breaking, 
which led to a chiral geometry 
of the inner symmetry crystal, as described in section 5. 

To be definite I will use the 7-dimensional 
vector product to describe the 
effective interaction among the phinon excitations 
after the phase transition to $A_4\times Z_2$, and I will show 
that it can be used to describe the required 
correlation between the chirality of the 
internal lattice and the handedness of the weak interaction 
in physical space. 

Before I proceed I want to review the known facts 
about the 7-dimensional vector product: 
just as in 3 dimensions a vector product can be defined 
in 7 dimensions which assigns to any two vectors $\vec v$ and $\vec w$ 
in $R^7$ a vector $v \times w$ perpendicular 
to $\vec v$ and $\vec w$ and of magnitude $|\vec v||\vec w|\sin(\angle)$. 

A cross product which fulfills these properties 
of bilinearity, orthogonality and magnitude exists 
in fact {\it only} in three and seven dimensions\cite{masi}, 
a statement, which is related to Hurwitz's
theorem, which says that normed division algebras 
can be defined only in 1, 2, 4 and 8 dimensions\cite{conway}. 

By definition, for the 7-dimensional vector product to work, 
one needs a seven-dimensional euclidean space. However, 
I will identify one of these dimensions with time, i.e. 
work in 6+1 dimensions with 3 dimensions being 
occupied by the internal lattice and the rest forming a 
3+1 Minkowski spacetime. The notation will be Euclidean 
throughout with an implicit Wick rotation ($x_0 \rightarrow i x_0$) 
understood, whereby part of the vector product terms 
appearing later e.g. in eq. (\ref{eq8abb}) become imaginary.

Unfortunately, the 7-dimensional cross product is 
less intuitive than the 3-dimensional one. 
For example, the choice of the perpendicular axis is not unique, 
which means that there are different possibilities to 
define the 7-dimensional cross product (which however 
turn out to be equivalent up to rotations of axes). 
One possibility to define is 
\begin{equation} 
\vec e_i \times \vec e_j =\epsilon_{ijk} \vec e_k
\label{eqaabb}
\end{equation}
where $\epsilon_{ijk}$ is a completely antisymmetric tensor with a
positive value +1 when ijk = 123, 145, 176, 246, 257, 347, 365\cite{masi}. 
Equivalently, one may use the rules given in table 2.
Assuming representations $\vec v=\sum_{i=1}^7 v_i \vec e_i$ 
and $\vec w=\sum_{i=1}^7 w_i \vec e_i$ 
one can then calculate e.g. the third component of the 
cross product of two vectors $\vec v$ and $\vec w$ as 
\begin{equation} 
(\vec v \times \vec w)_3=v_1 w_2 -v_2 w_1 +v_4 w_7-v_7 w_4 -v_5 w_6+v_6 w_5
\label{eq8abb}
\end{equation}

\begin{table}
\label{tabgg4}
\begin{center}
\begin{tabular}{|l|c|c|c|c|c|c|c|}
$\times$  &$\vec e_1$ &$\vec e_2$ &$\vec e_3$ &$\vec e_4$ &$\vec e_5$ &$\vec e_6$ &$\vec e_7$\\
\hline
$\vec e_1$   &0          &$\vec e_3$ &$-\vec e_2$&$\vec e_5$ &$-\vec e_4$&$-\vec e_7$&$\vec e_6$\\
$\vec e_2$   &$-\vec e_3$  &0 &$\vec e_1$&$\vec e_6$ &$\vec e_7$&$-\vec e_4$&$-\vec e_5$\\
$\vec e_3$   &$\vec e_2$  &$-\vec e_1$ &0&$\vec e_7$ &$-\vec e_6$&$\vec e_5$&$-\vec e_4$\\
$\vec e_4$   &$-\vec e_5$  &$-\vec e_6$ &$-\vec e_7$&0 &$\vec e_1$&$\vec e_2$&$\vec e_3$\\
$\vec e_5$   &$\vec e_4$  &$-\vec e_7$ &$\vec e_6$&$-\vec e_1$ &0&$-\vec e_3$&$\vec e_2$\\
$\vec e_6$   &$\vec e_7$  &$\vec e_4$ &$-\vec e_5$&$-\vec e_2$ &$\vec e_3$&0&$-\vec e_1$\\
$\vec e_7$   &$-\vec e_6$  &$\vec e_5$ &$\vec e_4$&$-\vec e_3$ &$-\vec e_2$&$\vec e_1$&0\\
\end{tabular}
\bigskip
\caption{Multiplication rules for the 7-dimensional vector product. 
The unit vectors $\vec e_i$ can be identified with the octonion units 
via $\vec e_1=I$, $\vec e_2=J$, $\vec e_3=IJ$, $\vec e_4=L$,
$\vec e_5=IL$, $\vec e_6=JL$, $\vec e_7=(IJ)L$ so as to fulfil 
the relation (\ref{eu79}).}
\end{center}
\end{table}

A related ambiguity is that for any cross product $\vec v \times \vec w$ 
in $R^7$ there are other planes than that spanned by v and w 
giving the direction of the vector $\vec v \times \vec w$. 
This can be seen from eq. (\ref{eqaabb}), where for any unit vector, 
which one chooses, there are three cross products of unit vectors, 
which takes its value (up to a sign).
Each of these cross products corresponds 
to another plane mapped into the given direction. 

To get a better understanding of rotations in higher dimensions, 
the notion of 'planes of rotation' is often introduced. 
A 'plane  of rotation' is a 2-dimensional linear subspace of a 
given d-dimensional space which is mapped to itself by a 
given rotation. 
In 3 dimensions it is the uniquely determined plane perpendicular 
to the axis of rotation, whereas  
for general dimension d there can be up to 
[d/2] planes of rotation for a given rotation, 
i.e. the maximum number of such planes is 1, 1, 2, 2, 3, 3 etc 
for dimensions 2,3,4,5,6,7 etc. 
If a rotation has several planes of rotation these are necessarily 
orthogonal to each other, with only the zero vector in common. 

All these statements rely on an elementary result of 
linear algebra, namely that any 
orthogonal transformation in $R^d$ can be 'diagonalized' 
to a set of 2-dimensional rotations by angles $\theta_i$, 
i=1,...,[d/2]
\begin{equation} 
O_2=
\left(\begin{array}{cc}
\cos(\theta_i) & \sin(\theta_i) \\
-\sin(\theta_i) & \cos(\theta_i)
\end{array}\right)
\end{equation} 
plus at least one unity on the diagonal of the matrix, if d is odd. 
The planes of rotation are the 2-dimensional subspaces, 
on which these 2-dimensional rotations are defined. 

Just as in 3 dimensions one may define a triple 
product as the scalar product of a vector with 
a cross product of 2 other vectors, cf. eq. (\ref{eq3334}). 
In 3 dimensions this pseudoscalar quantity 
gives the oriented volume of the parallelepiped 
spanned by the 3 vectors and is therefore invariant under 
the group SO(3) of proper rotations in 3 dimensions 
(but changes sign under parity). 
In 7 dimensions the situation is different\cite{cacci}:  
The invariance group is not full SO(7) 
but the exceptional Lie group $G_2 \subset$ SO(7), 
which comprises the $SO(3)\times SO(3)$ symmetry 
of inner and physical space (and also the 
point group symmetries of the corresponding lattices).  

There is an interesting and illuminating relation of the 
cross products in 3 and 7 dimensions with quaternions and 
octonions, respectively. They can in fact be 
related to the imaginary part of the product of 
two quaternions or octonions V and W\cite{conway}.  
To see this one should identify the basis $\vec e_i$, $i=1,...,7$ of $R^7$ 
as used in eq. (\ref{eqaabb}) with the octonion units 
$I,J,IJ,L,IL,JL,(IJ)L$ and then show that the cross product of two 7-dimensional 
vectors is given by 
\begin{equation}
\vec v \times \vec w = \frac{1}{2}Im (VW-WV)
\label{eu79}
\end{equation}
where on the right hand side octonion multiplication has been 
used on the imaginary octonions V and W,  
$V=v_1 I +v_2 J+v_3 IJ +v_4 L +v_5 IL +v_6 JL + v_7 (IJ)L$ 
(and similar for W).
Note that a change in sign in the definition of the 
associator $(IJ)L=-I(JL)$ would reverse all PV effects. 


At this point the reader may complain, that after all 
these elaborations there is still no real progress in 
interrelating internal and weak parity effects. 
After all, the rotations within two different 
planes of rotation, one internal and 
one in physical space, a priori are completely independent. 
However, we shall shortly see that the rules of the 
7-dimensional cross and triple products given in table 2 
establish a connection between the rotary directions. 

To this end I come back to the concrete example of production 
and decay of a W-boson eq. (\ref{eq2279g}). As before it is considered 
in the rest frame of the W, and with relativistic fermions. 
In the present 3+3+1 dimensional model let 
$P_u=(p_u,\vec q_u)$ be the 7-momentum of the up-quark 
with $p_u$ given in eq. (\ref{eq2880g}) and $\vec q_u$ 
the crystal momentum of the corresponding phinon in internal space, 
and similarly for the other particle momenta. 

Concerning the parity issue the W-boson has two 
relevant planes of rotation. 
One is spanned by $\vec e_2$ and $\vec e_3$ and 
given by the left handed W-polarization eq. (\ref{eq2280g}) 
which enters via the factor 
\begin{equation} 
\label{eq6011335}
-i(\vec\epsilon_L \times \vec\epsilon_L^\ast)
=\vec e_2 \times \vec e_3=\vec e_1 
\end{equation}
in eq. (\ref{eq22gh1}).
The other one is the plane of rotation of the 
internal helix which any phinon mode has to respect
and which in general will be left handed and spanned by two vectors 
$\vec g$ and $\vec h$. When the $W^+$ is produced, they 
may be fixed without loss of generality to be given 
by $\vec g=\vec e_5$ and $\vec h=\vec e_4$ so that 
\begin{equation} 
\label{eq601141}
\vec g \times \vec h=-\vec e_1 
\end{equation}
according to table 2.

The situation may be compared to that in 
optical activity of chiral molecules or crystals, 
where the polarized photon induces a current in the 
chiral geometry, which in turn influences the photon 
in such a way that its polarization is rotated. 
Geometrically this is due to the interplay of two chiral structures: 
one from the polarized photon, and the other one from the chiral 
object. 

In the present case the two chiral objects a priori are 
independent, because they live in two different subspaces 
of the 6+1 dimensional world. 
The plane spanned by $\vec g$ and $\vec h$ as well as the 
crystal-momenta of the phinon modes $\vec q_u=-\vec q_d$ are 
all orthogonal to physical space. 
However, the rules of the 7-dimensional vector product 
in table 2 are such that they give $\vec g\times \vec h=
\vec e_5 \times \vec e_4=-\vec e_1$, 
i.e. they put the internal helical structure in parallel 
with the external angular momentum 
$i(\vec\epsilon_L \times \vec\epsilon_L^\ast)=-\vec e_1$ 
and in general forbid $\vec g\times \vec h$ to point 
into one of the inner directions.
It is only due to this peculiar structure of the 7-dimensional vector product 
that the two chiralities which in principle live in two 
different subspaces can meet in the form of an interaction 
which ties them together. 
In fact, when the $W^+$ decays to $\mu^+(P_l) + \nu_\mu(P_n)$, the 
plane of the internal helix will become 
directed towards the neutrino momentum $\vec g \times \vec h=-\hat p_n$, 
and the parity violating term 
in the Standard Model prediction eq. (\ref{eq22gh1}) can be 
reconstructed from a matrix element, which is the product 
of an internal contribution $<q_W|S_{in}|q_l q_n>$ 
and a spatial contribution $<p_W|S_{sp}|p_l p_n>$, as 
\begin{equation} 
[-i(\vec \epsilon_L\times \vec \epsilon_L^\ast)\hat p_n]
 [(\vec g\times \vec h)\hat p_n]
=-i(\vec \epsilon_L\times \vec \epsilon_L^\ast)(\vec g\times \vec h)
=\vec e_1 \hat p_n 
\label{eq6011}
\end{equation}

The interaction given by the 7-dimensional vector product is antisymmetric under 
odd permutations in both its internal and physical part. 
This is precisely what is needed to describe the parity violation, 
simply because one will get negative 
energy contributions to the partition function, if an odd 
transformation is applied to the points of a tetrahedron. 

The 7-dimensional triple product will thus give a good description of 
the parity violating effect. 
However, it should be kept in mind, that this 
description is only an effective one, because  
it is originally induced by the chiral geometric 
structure of the internal lattice. The geometrical 
structure comes first, so that when two phinons 
collide they have to respect the chirality 
of the $A_4$ symmetry in such a way that only 
spatially lefthanded states can enter the game. 
What the 7-dimensional triple product does, is to 
to connect the parity effects in internal and physical 
space in such a way that the chiral nature of the inner 
symmetry lattice induces parity violation in physical space.

\section{The Mignon Alternative}

There is another interpretation of the tetrahedral ordering 
of the spectrum eq. (\ref{eq8hg}), namely by using the 
rotational instead of the translational transformation properties 
in the last column of table 1. 
In that case one is led to consider spin models, 
and the vector displacements $\vec u$ of the atomic vibrations 
are replaced by axial vectors $\vec S$ 
which fulfil angular momentum commutation relations. 

Spin models have been 
considered in statistical and solid state physics for a 
long time\cite{spinm1,spinm2,wen1}, and they have been used to describe 
magnetic phase transitions and excitations as well as many other phenomena. 
In contrast to ordinary spin models, which were first used 
by Heisenberg to describe ferromagnetic effects in solids, 
the spin models in this section are defined on inner 
symmetry space, because I want to use them 
as an alternative to the phinon picture 
developed in the preceeding sections. 
Phinons are then to be replaced by 'mignons', 
quasi-particle excitations which differ from 
magnons by living on the internal $A_4\times Z_2$ 
crystal structure, i.e. they are defined with the help of 
an internal 3-dimensional 'spin vector' $\vec S(m)$ 
given on lattice points m and 
interacting via a Heisenberg like internal Hamiltonian 
\begin{equation}
H_H = - \sum_{m,m'} J(m,m') \vec S (m) \vec S(m')
\label{eq1777}
\end{equation}
where $J(m,m')$ are the coupling strengths and 
the sum runs over all neigbouring lattice sites. 
Depending on the sign of J one has ferromagnetic- or antiferromagnetc 
type of interactions.
In contrast to phinons, which arise 
from vibrational displacement vectors $\vec u$, 
the spin vectors are (internal) axial vectors and 
have (internal) angular momentum properties, as will be shown below.

The physics behind such a model relies on exchange 
integrals of fermion wave functions $\psi_{\pm \frac{1}{2}}$ 
defined on the inner crystal of the form 
\begin{equation} 
I_E=\frac{1}{2} \int d^3r_1 d^3r_2 \sum_{\alpha,\beta=1}^2 
\psi_\alpha^\dagger(\vec r_1) \psi_\beta^\dagger (\vec r_2) V(|\vec r_1 - \vec r_2|) 
\psi_\beta (\vec r_1) \psi_\alpha (\vec r_2)
\label{em5}
\end{equation}
where the sums run over 2 internal 'spin' directions $\alpha, \beta =\pm \frac{1}{2}$, 
V is an internal potential and $\vec r_i$ are vectors in internal space. 
These exchange integrals can be shown to lead directly to the 
Heisenberg Hamiltonian eq. (\ref{eq1777}) while other 
possible contributions which may look interesting for chiral 
crystal structures like the 
the Dzyaloshinsky-Moriya interactions \cite{dzy1,dzy2} 
$\vec D(m,m') [\vec S (m) \times \vec S(m')]$
are not generated in leading order. 
They can arise e.g. by spin-orbit corrections and are 
usually suppressed. 

The following relation constitutes the link between 
the internal spin vector $\vec S$ in 
eq. (\ref{eq1777}) and the internal spinor $\psi$ in eq. (\ref{em5}): 
\begin{equation}
\vec S = \frac{1}{2} \psi^\dagger \vec \tau \psi 
\label{eq557}
\end{equation}
where $\vec\tau$ is the triplet of internal Pauli matrices. 

The possibility of a 6+1 dimensional discrete spacetime 
which soon after the big bang got split to 3+3+1 dimensions 
has been repeatedly discussed in the previous sections, 
and I will in fact assume in the following that the original 
dynamical field is a (6+1)-dimensional spinor, 
whose 'mignon' excitations on the inner symmetry space 
can be identified with the quark, lepton and gauge boson 
fields appearing in the Standard Model. 

Such a spinor can be considered a SO(7) spinor 
by a Wick rotation\footnote{More generally, 
in $SO(d_1,d_2)$ the spinor dimensions viewed over complex space 
coincide with the case of the $(d_1+d_2)$-dimensional Euclidean space.}
and to two SO(6) spinors (one for a SO(6) fermion and the other 
for an antifermion) in the non-relativistic limit. 
In general, higher dimensional spinors can be 
constructed from building blocks of spinors in 
lower dimensions, as described for example in ref. \cite{rossbook}. 
In the case at hand, 
they have a close relationship to the octonion 
algebra \cite{dixonbook}, just as 
3-dimensional spinors have to quaternions. 
While the Pauli matrices can be identified 
more or less directly with the quaternion units, the situation 
in 7 or 6+1 dimensions is somewhat more subtle 
because the octonion algebra is not associative, i.e. the 
octonion units I, J, IJ, L, IL, JL and (IJ)L 
cannot be exactly represented by 7-dimensional matrices. 
A detailed and explicit description of the relationship 
can be found, for example, in the book by Dixon \cite{dixonbook}.

While the covering group of SO(6) is isomorphic 
to SU(4) and has 2 fundamental complex representations 
$\underline{\boldsymbol{4}}$ and $\underline{\boldsymbol{4}}^{\ast}$ 
which are conjugate to each other, 
the covering group of SO(7) is Spin(7) 
with one spinor representation of dimension $\underline{\boldsymbol{8}}$, 
which is again related to the non-associative 
division algebra of octonions\cite{dixonbook,conway,kantor,schaf}. 
Breaking Spin(7)$\rightarrow$SU(4) 
there is a decomposition $\underline{\boldsymbol{8}} \rightarrow 
\underline{\boldsymbol{4}} + \underline{\boldsymbol{4}}^{\ast}$ 
which reveals the particle antiparticle content of the 
original SO(6,1) spinor\cite{rossbook}. 

For simplicity let me concentrate on the nonrelativistic SO(6) fermion. 
The covering group of SO(6) is SU(4), $\widetilde{SO(6)}=SU(4)$. 
Dividing it into a spatial and an inner symmetry part 
$SU(2)^{sp}\times SU(2)^{in}$ 
reduces its spinor representation 
$\underline{\boldsymbol{4}} \rightarrow 
(\underline{\boldsymbol{2}^{sp}},\underline{\boldsymbol{2}^{in}})$ further 
to a field $\phi_{a,\alpha}$ 
which is a doublet both under $SO(3)^{sp}$ and under $SO(3)^{in}$, 
i.e. the index a acts as a Pauli spinor index in physical space
whereas $\alpha$ is a 'spinor' index in the inner symmetry space. 
Going to an inner lattice with $A_4$ symmetry, i.e. 
a space with symmetry group $SO(3)^{sp}\times A_4^{in}$, 
one obtains states which transform as 
$(\underline{\boldsymbol{2}^{sp}},G_1^{in})$ 
where $G_1$ is the 2-dimensional spinor representation of $A_4$ 
obtained by restricting the $\underline{\boldsymbol{2}^{in}}$ 
representation of $SU(2)^{in}=\widetilde{SO(3)^{in}}$ to $A_4^{in}$. 
If physical space has a lattice structure, too, one 
ends up with $(G_1^{sp},G_1^{in})$. 
The overall group theoretic situation is summarized in table 3. 

\begin{table}
\label{tab2aa}
\begin{center}
\begin{tabular}{|c|c|c|}
\hline
& & \\
the fundamental fermion& 
$Spin(7) \leftrightarrow \widetilde{SO(6,1)}$ &
$\underline{\boldsymbol{8}}=F$ \\ 
& & \\
\hline
& & \\
nonrelativistic limit& 
$\widetilde{SO(6,1)} \rightarrow SU(4)$ & 
$\underline{\boldsymbol{8}} \rightarrow \underline{\boldsymbol{4}} 
+ \underline{\boldsymbol{4}}^{\ast}$ \\ 
& & \\
\hline
& & \\
inner and physical space split&
$SU(4) \rightarrow SU(2)\times SU(2)$ &
$\underline{\boldsymbol{4}} \rightarrow (\boldsymbol{2_{in}},\boldsymbol{2_{sp}})
     =\phi_{a,\alpha}$ \\
& & \\
\hline
& & \\ 
restriction to a lattice &
$SU(2)\times SU(2) \rightarrow 
    \widetilde{S_4^{sp}} \times \widetilde{S_4^{in}}$ &
$\underline{\boldsymbol{4}} \rightarrow (G_1^{sp},G_1^{in})$ \\
& & \\
\hline
\end{tabular}
\bigskip
\caption{Group theoretic view on the fundamental dynamical 
field F in the tetron model. The tilde denotes covering groups.}
\end{center}
\end{table}

We now make the further assumption that the compactification 
process completely separates the interactions 
in physical and internal space. Then the 
eigenfunctions $\phi_{a,\alpha}$ will factorize as 
$\phi_{a,\alpha}=f_a \psi_\alpha$ where $\psi_\alpha$ 
constitutes the internal spin vector $\vec S$ 
according to eq. (\ref{eq557}), 
which will follow the dynamics of the inner Heisenberg 
model eq. (\ref{eq1777}), while the physical spinor part $f_a$ (or its 
relativistic extension) fulfills the Dirac equation. 
The corresponding Bloch waves are a product of an internal spin wave (='mignon') 
and a free Dirac wave 
\begin{equation} 
\Psi_{ia}= 
f_a(\vec x) S_i(\vec r) 
e^{i(\vec p \vec x - \nu t)} 
e^{i(\vec q \vec r - \omega t)}
\label{ex4hg}
\end{equation}
i.e. at a given point the free Dirac wave 
must be multiplied by the mignon mode. 

If given on different spots $m$ and $m'$ the inner 
spin vectors commute; however given on the same spot $m$ they 
fulfil angular momentum commutation relations 
\begin{equation} 
[S_i (m), S_j (m)] = \epsilon_{ijk} S_k (m)
\label{em7}
\end{equation}
where $i,j,k \in {1,2,3}$ denote the 3 directions in the internal crystal. 

The equations of motion for the operators $\vec S (m)$ 
can be obtained most easily in the Heisenberg picture: 
\begin{equation} 
\frac{d \vec S (m)}{dt}=  \frac{i}{\hbar} [H_H,\vec S (m)]
\label{em8}
\end{equation}
Inserting $H_H$ from eq. (\ref{eq1777}) leads to 
\begin{equation} 
\frac{d  S_i (m)}{dt}=  -\epsilon_{ijk} \sum_{m'} J(m,m') 
                          S_j(m) S_k(m')
\label{em9}
\end{equation}

These equations decouple, if one 
defines $S_\pm (m) = S_1 (m) \pm iS_2 (m)$, 
and one can show that 
there are plain wave solutions of the form 
\begin{equation} 
S_\pm (m) =  S_\pm(\vec q) e^{i (\vec q \vec r (m)-\omega t)}        
\label{em10}
\end{equation}
which are called internal spin waves or mignons. 
More precisely, the x- and y-components turn out to be out of phase by 
an angle $\pi/2$, so that $S_1(t)$ and $S_2(t)$ can be said to precess on a cone.

Just as for phinons the next step is to consider 
nonlinear effects which lead to interactions among mignons, 
to scattering processes and possibly to bound states. 
For the following discussion it is convenient to switch 
to the language of the second quantization. 
We could have done so for phinons as well, however 
at will, whereas here it is really advisable. 
The point is, that as bilinears of internal fermions 
the spin operators $S_i$ and the associated quasi-particles 
should be internal bosons. 
However, as they stand they do not fulfil the 
canonical commutation relations for bosons, 
and this can be remedied most easily in the framework 
of the second quantization.  
Namely, in order to recover the ordinary boson commutation 
relations, they may be rewritten 
with the help of a Holstein-Primakoff transformation\cite{holstein} as 
\begin{eqnarray} 
S_+(m) &=& \sqrt{2S-a^\dagger(m) a(m)} a(m)          \\ 
S_-(m) &=& a^\dagger(m) \sqrt{2S-a^\dagger(m) a(m)}  \\ 
S_z(m) &=& S - a^\dagger(m) a(m)
\label{em12}
\end{eqnarray}
One can prove that the commutation relations eqs. (\ref{em7})
are satisfied for this representations iff the 
creation and annihilation operators $a^\dagger(m)$ and $a(m)$ 
fulfil the canonical commutation relations for harmonic oscillators. 

The spin S in eq. (\ref{em12}) is defined by $\vec S(m) \vec S(m)=S(S+1)\mathbf{1}$ 
and the products $a^\dagger(m) a(m)$ are number operators 
with eigenvalues $0,1,2,...$. Since $S_z(m)$ is supposed 
to take eigenvalues $-S,...S-1,S$ one has to impose the 
constraint $a^\dagger(m) a(m)\le 2S$.  
This inconvenience is compensated for by the fact that 
the vacuum of the Holstein-Primakoff bosons fulfills\cite{holstein} 
\begin{eqnarray} 
a^\dagger(m) a(m) |0(m)> = 0 \quad \Longleftrightarrow \quad S_z(m)|0(m)>=S|0(m)>
\label{em13aa}
\end{eqnarray}
and can thus be identified with the 'ferromagnetic' ground state. 


For practical calculations the roots appearing in eq. (\ref{em12}) 
should be expanded in powers $a^\dagger(m) a(m)/2S$. 
Using momentum space creation and annihilation operators 
\begin{eqnarray} 
a(m)&=& \sum_{\vec q} \exp [i\vec q \vec r(m)] 
b(\vec q) \nonumber \\
a^\dagger (m)&=& \sum_{\vec q} \exp [-i\vec q \vec r(m)] 
                      b^\dagger(\vec q)
\label{em13}
\end{eqnarray}
the Hamiltonian of the system can be expressed as 
\begin{equation} 
H_H=E_0 + \sum_{\vec q} \hbar \omega(\vec q) b^\dagger(\vec q) b(\vec q) 
+\sum_{\vec q,\vec q'} V(\vec q, \vec q') b^\dagger(\vec q')
b^\dagger(-\vec q') b(\vec q) b(-\vec q)
\label{em15}
\end{equation}
where $E_0$ is the ground state energy and 
it has been assumed that stable mignon pairs exist 
only if the two mignons have opposite wave vectors. 

Turning now to the special case of the $A_4\times Z_2$ lattice 
under discussion in this paper, one may approximately introduce 
two coupling constants, called J and J', one which couples 
atoms within the lattice tetrahedrons and one which couples two 
tetrahedrons (or equivalently a $Z_2$-pair). 
As in section 3 one may examine the limiting cases of large J and large J', 
respectively, and perform a calculation within the mean field approximation. 
For that one should use appropriate local states, dublet states 
$\frac{1}{2}(|\uparrow> \pm |\downarrow>)$ for the $Z_2$ excitations 
and 16 states built from $|\uparrow\uparrow\uparrow\uparrow>$, 
$|\uparrow\uparrow\uparrow\downarrow>$ etc for a tetrahedron. 

Within this framework the following 
bilinears can be replaced by their thermal averages: 
\begin{eqnarray} 
b(\vec q) b(-\vec q) &=& <b(\vec q) b(-\vec q)> \nonumber \\
b^\dagger(\vec q) b^\dagger(-\vec q) &=& <b^\dagger(\vec q) b^\dagger(-\vec q)> 
\label{em18}
\end{eqnarray}
so that the interaction part of the Hamiltonian can be approximated by 
\begin{eqnarray} 
H_H &=& E_0+\frac{1}{2} \sum_{\vec q} \hbar \omega(\vec q) 
[b^\dagger(\vec q) b(\vec q) + b^\dagger(-\vec q) b(-\vec q)] \nonumber \\
& & +\sum_{\vec q,\vec q'} V(\vec q, \vec q') 
[ b^\dagger(\vec q') b^\dagger(-\vec q') 
<b^\dagger(\vec q) b^\dagger(-\vec q)> \nonumber \\
& &+ b(\vec q)  b(-\vec q) <b^\dagger(\vec q') b^\dagger(-\vec q')>
- <b(\vec q)  b(-\vec q)> <b^\dagger(\vec q') b^\dagger(-\vec q')>]
\label{em19}
\end{eqnarray}

This result may be used as a starting point to analyze 
phase transitions and mignon-mignon scattering processes, 
very similar to eq. (\ref{eq33332g}) for phinons. 
For example, using shifted frequencies 
\begin{eqnarray} 
\Omega(\vec q) &=& \omega(\vec q) [1+\frac{E_0}{8} \sum_{\vec q'} \hbar \omega(\vec q')
<b(\vec q) b(-\vec q)>] 
\label{em21}
\end{eqnarray}
one may put $H_H$ into the form\cite{zhang} 
\begin{eqnarray} 
H_H &=& E_0 +\frac{1}{2} \sum_{\vec q} \hbar \Omega(\vec q) 
[b^\dagger(\vec q) b(\vec q) + b^\dagger(-\vec q) b(-\vec q)] 
\label{em20}
\end{eqnarray}


\section{Conclusions}


In the present paper the tetron model of elementary particles has been 
further developed, and arguments have been given how some old puzzles of 
high energy physics can be explained by giving a discrete 
structure to a 3-dimensional inner symmetry space, which has 
the form of a {\it finite internal crystal lattice} 
with 8 points in its unit cell. 

The fermions and gauge bosons of the Standard Model have 
been interpreted as quasi-particle excitations, 
which arise on this lattice from vibrations or quantum exchange 
of more fundamental entities. 
Since the phenomenological indications are not conclusive, 
I have analyzed two kinds of models, one describing 
vibrational excitations on the internal space (called 'phinons'), 
the other one with 'mignons'.  
Bound states from these excitations have been examined 
as well as symmetry breaking scenarios, which may have led 
to phase transitions in the early universe. 

It is an old dream of theoretical physicists 
that inner symmetries may be obtained by extending 
ordinary space to higher dimensions (see for example 
ref. \cite{kaluza}). 
The present paper shows how the connection between inner and 
outer symmetries has to look like 
in order to handle the observed parity 
violation of the weak interactions. 
It is argued that at the big bang the universe 
started with a unified discrete 6+1 dimensional spacetime. 
Afterwards it underwent at least 2 phase transitions: 
in the first it split into a physical and a compact internal part, 
where the latter cannot be directly perceived, 
but is only observable by effect of charges and interactions. 
At a later stage, when physical and internal space 
cooled down further towards the weak interaction scale, 
there was a phase transition within the internal lattice, 
which led to a chiral structure with point 
group $A_4\times Z_2$. 


As explained in footnote 1, 
I have not been definite about using a discrete base 
spacetime or not. However, if one is willing to make 
this additional assumption, one encounters 
advantages as well as drawbacks. The advantages are that 
ultraviolet divergences are regularized per se and no
renormalization is needed when it comes to calculations 
involving short distances. Furthermore, no-go theorems 
like the Weinberg-Witten theorem \cite{weinberg} which in the
continuum forbid the unification of spatial and inner symmetries do not apply.
Finally, of course, the lattice ansatz naturally explains the selection 
rule mentioned in ref. \cite{lampe1} that all physical states must be 
permutation states: just because the lattice excitations 
must transform under representations of its point group $A_4\times Z_2$. 

A credit point not mentioned at all in this paper 
is that a lattice ansatz, which not only discretizes 
internal but also physical space, 
naturally explains why there are no
fundamental particles with spin larger than 2 
in nature: this has to do with the fact 
that the discrete symmetry groups under discussion 
possess only irreducible representations with small dimensions. 
For example, assuming a cubic spatial lattice with $S_4^{sp}$ 
as a point group, there are (ray)-representations of dimension 
1 ($A_{1},A_{2}$), 2 ($G_1,G_2,E$), 3 ($T_1,T_2$), and 4 ($H$). 
$G_1$ has been discussed as the relevant spatial spin-1/2 
representation for fermions in the last section, but 
one may justly ask the question, 
why the other modes, like e.g. the spin-3/2 
representation $G_2+H$ are not observed. 
This question 
has been addressed in a separate publication\cite{lampec}, 
where those modes were suggested as dark matter candidates. 

Now for the drawbacks: first of all, on a spatial lattice Lorentz symmetry is 
broken, which is of course in contradiction to everyday experience. 
However, for a classical observer Lorentz symmetry can be restored 
by assuming the lattice spacing to be either unobservably small 
or that it may be a fluctuating quantum lattice, 
where the lattice points move around randomly and 
follow some quantum stochastic 
process \cite{santiago}. There is some relation of this idea 
to other models which involve a fundamental 
length scale, like quantum foam models \cite{wheeler,amb,yam}, 
which however assume gravity to play the 
central role in the dynamics, 
while in the present model gravitational interactions 
and cosmological phenomena 
appear only as byproducts of the tetron lattice 
interactions, as shown in ref. \cite{lampec}.
 
Secondly, it is usually difficult to define fermions on a lattice without 
getting problems with (micro)causality, because in contrast to 
bosons fermions 'know' about other fermions on neighbouring 
lattice sites and this induces nonlocal correlations 
and possible synchronisations beyond the event horizon. 
Fortunately, this is not the case in the approach presented here, 
because the fundamental fermion field introduced 
in the last section is defined within the whole crystal 
(both for internal and physical space) not just on the 
lattice points.

I have given arguments how the gauge 
structure and also the parity violation of the 
weak interaction have arisen from properties 
and interactions within the internal crystal. 
At the present stage, the true nature of the underlying dynamics 
that controls the vibrational or 'magnetic' excitations remains 
unclear. For example, it is possible that it 
turns out to be in some sense supersymmetric.  
However, one should consider 
this option far from being compelling. 
In particular, the appearance of discrete lattice structures 
above Planck distances gives no indication that 
the fundamental Lagrangian will have anything to do 
with the nowadays popular superstring or M-theories.

\end{document}